\documentclass[
]{ceurart}
\usepackage{listings}
\usepackage[utf8]{inputenc}

\begin{document}

\copyrightyear{2022}
\copyrightclause{Copyright for this paper by its authors.
  Use permitted under Creative Commons License Attribution 4.0
  International (CC BY 4.0).}

\conference{PoEM’2022 Workshops and Models at Work Papers, November 23-25, 2022, London, UK}

\title{Executable Models and Instance Tracking for Decentralized Applications - Towards an Architecture Based on Blockchains and Cloud Platforms}


\author[1]{Felix Härer}[%
orcid=0000-0002-2768-2342,
email=felix.haerer@unifr.ch,
url=https://www.unifr.ch/inf/digits/en/group/team/haerer.html,
]
\address[1]{Digitalization and Information Systems Group, University of Fribourg, Bd de Pérolles 90, 1700  Fribourg, Switzerland}

\begin{abstract}
The execution of decentralized applications on blockchains is limited today by technical and organizational barriers, including scalability and the high complexity to specify execution correctly for developers as well as for domain experts in organizations. Overcoming these limitations could allow for decentralized coordination beyond data, where distributed parties rely on higher-level abstractions for coordinating their actions using decentralized applications, not limited to organizations. Towards this goal, the paper at hand proposes executable models as high-level abstraction that can be observed and tracked by distributed parties. In particular, it is investigated how executable models on cloud platforms can be coupled with smart contracts for tracking their execution, concluding with an architecture as exploratory research result towards supporting scalability and decentralized coordination.
\end{abstract}

\begin{keywords}
  Blockchain \sep
  Decentralized Applications \sep
  Executable Models \sep
  Instance Tracking \sep
  Serverless Computing
\end{keywords}

\maketitle

\section{Introduction}

In recent years, blockchains have introduced novel technical infrastructures for exchanging transactions and executing applications online. They establish ledgers for transactions~\cite{nakamotoBitcoinPeertoPeerElectronic2008,buterin_ethereum:_2013,diAngeloTokens2020}, e.g. using tokens as a unit of account for payments or goods~\cite{diAngeloTokens2020}, and allow involved parties to track and validate transactions without trusted third parties. In effect, end users or businesses can interact through verified transactions and track the flow of tokens by attaching them to goods or any physical or digital item without centralized coordination. These scenarios are enabled by \textit{decentralized applications} executing \textit{smart contract} programs on blockchains, where data processing and transactions are verifiably executed according to the source code~\cite{antonopoulosMasteringEthereumBuilding2019}. While blockchains enable verifiable transactions and execution, there remain technical and organizational barriers for the design of decentralized applications.

In particular, current challenges concern scalability on a technical level, limited by the processing capabilities and consensus algorithms of today's blockchains when also pursuing security and decentralization~\cite{ZhouScalability2020}. Secondly, for organizational adoption, domain experts are essential in operations and would be involved in designing decentralized applications in addition to developers. Thus, the technically complex development of smart contracts becomes even more challenging in an organizational context, given that technical implementations need to execute correctly according to business, domain, and technical requirements. 

Model-based representations could address these challenges when applied for the execution of decentralized applications. Using executable models as higher-level representations, distributed parties could potentially coordinate their actions using decentralized applications. Not limited to organizations, models could provide more formality, meaningful and modular abstractions, and interactivity, as demonstrated recently in low-code and no-code development~\cite{curty_blockchain_2022}. 

Towards applying executable models for decentralized coordination, the paper at hand pursues the research question of how executable models and their instances can be coupled with decentralized applications using smart contracts. It builds on an existing concept for instance tracking~\cite{harer_decentralized_2018, harer_integrierte_2019, harer_process_2020} and contributes a dedicated architecture for executable models on cloud platforms coupled with smart contracts for tracking execution. Through exploratory research, this architecture has been constructed for scalability while allowing distributed parties to track execution steps for coordinating their actions without centralized entities.

The remainder of this paper is structured as follows. Section 2 introduces related work, identifying characteristics missing in present approaches. Section 3 introduces execution and instance tracking concepts and concludes with an architecture and smart contract implementation. Section 4 concludes.

 
\section{Related Work}

This section introduces related work on blockchains, decentralized coordination, and their utilization in model-based approaches.

Blockchains are a class of technologies permitting the decentralized storage and processing of data. Blockchain platforms such as Bitcoin~\cite{nakamotoBitcoinPeertoPeerElectronic2008} and Ethereum~\cite{buterin_ethereum:_2013} achieve decentralization, the coordination without a centralized entity~\cite{simonCentralizationVsDecentralization1954,harer_process_2020}, through data structures verifiable for integrity and distributed nodes in a network storing, processing, and verifying data through consensus algorithms. Decentralized applications are executed, at least in part, through data processing on the blockchain, e.g. using smart contracts interpreted by the Ethereum Virtual Machine~\cite{antonopoulosMasteringEthereumBuilding2019}. 

On this basis, decentralized coordination concepts beyond the level of data are explored. Decentralized autonomous organizations (DAO)~\cite{santana_blockchain_2022} are an example of realizing governance by their participants on a technical level, e.g. through voting mechanisms in smart contracts. On a business level, the concept of DAOs has been applied for establishing decentralization in businesses in a model-based approach~\cite{harer_decentralized_2018, harer_integrierte_2019, harer_process_2020}. It supports designing business systems with processes and the tracking of their instances; however, not involving execution on the blockchain. Business process execution has been introduced before in~\cite{weberUntrustedBusinessProcess2016}; however, not involving the design of business systems or processes on the blockchain. Generally, business process management (BPM) using blockchains has since been recognized~\cite{mendling_blockchains_2018} and implemented primarily for business process monitoring and execution~\cite{garcia-garcia_using_2020}. Implementations rely often on model-based approaches, such as Lorikeet and Caterpillar ~\cite{tranLorikeetModelDrivenEngineering2018,lopez-pintadoCaterpillarBusinessProcess2019} using the BPMN 2.0 modeling language~\cite{omgBusinessProcessModel2014}. Further approaches subsequently followed, primarily for interorganizational and collaborative processes~\cite{falazi_modeling_2019, corradini_model-driven_2021, loukil_decentralized_2021}. In enterprise modeling and conceptual modeling, a custom blockchain for models in enterprises with permission mechanisms~\cite{fillKnowledgeBlockchainsApplying2018} has been suggested as well as an approach for the attestation of conceptual models on the Ethereum blockchain~\cite{harerDecentralizedAttestationConceptual2019b}.

Concerning model-based execution approaches, existing works support this concept on the blockchain in combination with execution engines. Notably, the Camunda workflow engine for BPMN 2.0~\cite{lopez-pintadoCaterpillarBusinessProcess2019} and state machine implementations~\cite{jurgelaitis_solidity_2022} have been proposed. While these approaches build on specific languages, the model-based execution could also be generalized and coupled with execution engines on cloud platforms, e.g. supporting serverless computing~\cite{yussupov_serverless_2020}, for combining decentralization and scalability. The paper at hand proposes an architecture towards this vision through the concept of instance tracking.

\section{Instance Tracking Architecture for Executable Models}

For the instance tracking of executable models, this section introduces the relevant concepts, establishes an architecture, and demonstrates a smart contract implementation.

\subsection{Concepts}

The concepts for executable models and instance tracking are introduced first, leading to the architecture.

\subsubsection{Executable Models}

Executable models are the basis and primary representation for users, e.g. end users or businesses, to design and specify execution for decentralized applications. In contrast to specifying source code, executable models are used as an abstraction representing the execution specification. The relevant concepts encompass:

\begin{itemize}
    \item \emph{Executable Models.} An executable model is a representation that abstracts from the specification by source code and can be executed by software on execution engines. Examples are state machines~\cite{amazon_2022} or executable BPMN models~\cite{omgBusinessProcessModel2014}.
    \item \emph{Execution Engines.} Execution engines support standardized modeling languages, e.g. the Camunda Workflow Engine for executing BPMN 2.0 models~\cite{camunda_2022}, or domain-specific models such as specialized state machine models, e.g. on cloud platforms such as AWS~\cite{amazon_2022}. In addition, modeling tools might provide an execution engine, e.g. AnyLogic 8 for modeling simulations~\cite{borshchev_big_2020}. For decentralized coordination, the execution engine must support distribution and scalability.
    \item \emph{Decentralized Applications.} Applications relying on non-centralized technical infrastructures and coordination principles can be considered decentralized applications~\cite{harer_integrierte_2019}. Non-centralized technical infrastructures operate abstract from individual infrastructure components that could represent single points of failure, while non-centralized coordination concerns operating without trusted third parties, e.g. by sharing responsibilities and control over development, usage, and governance aspects.
\end{itemize}

For example, decentralized web applications might realize non-centralized technical infrastructures by distributed file systems~\cite{psarasInterPlanetaryFileSystem2020}, blockchains, or cloud services where programs or functions are specified independent of servers or virtual machines by serverless computing~\cite{amazon_2022}. Non-centralized coordination can be achieved, e.g., through smart contracts, on-chain or off-chain voting, or role-based access control.


In combination of the concepts, decentralized applications can be based on executable models that utilize cloud-based execution engines together with blockchain and smart contract components. In this way, distributed users rely on a shared representation of the execution.

\subsubsection{Instance Tracking} 

Instance tracking is a concept for monitoring the execution of business process models by decentralized users on a blockchain~\cite{harer_decentralized_2018,harer_process_2020} that can be generalized and applied to decentralized applications.

\begin{itemize}
    \item \emph{Instance Tracking.} In terms of conceptual modeling~\cite{Sinz19}, instances are manifestations of models at run-time, representing objects or data of the abstract concepts defined by a model. Instance tracking refers to the observation of instances over time, either by recording discrete states of instances at run-time, or by recording state changes. 
    \item \emph{Instance Protocol.} An instance protocol records for a specific model the creation of instances, the discrete states or state changes for each instance, and the termination of instances. In the case of discrete states, initial, expected, and final states can be defined if relevant for the domain. In the case of state changes, a state transfer function or operations resulting in state changes can be defined. Additionally, the protocol might store metadata, such as involved users or blockchain addresses.
\end{itemize}

By registering models and tracking their instances with a smart contract over time, blockchain-based instance protocols can be created. Distributed users can track state changes and verify the consistency to the models, allowing for decentralized coordination using models and instance-level data, and data analysis for distributed applications. While this approach is not limited to specific application areas, it could be applied to track the state of, e.g., decentralized organizations or processes~\cite{harer_process_2020}, tokens in financial applications, marketplaces or supply chains, as well as data-intensive applications such as for distributed IoT devices, vehicles, industrial machines, or distributed sensor data.

\subsection{Architecture}

This section establishes an architecture for the decentralized instance tracking of executable models by identifying required components, discussing their behavior, and instance protocols for tracking and verifying relevant execution steps. 

The following requirements for components of a dedicated architecture address instance tracking by blockchains, adapted from ~\cite{harer_decentralized_2018, harer_process_2020}, and execution according to the respective concepts.

\begin{enumerate}
    \item \emph{Recording of models and instances}: Design aspects of an execution represented in models must be distinguished from instances and their states and transitions. While instances might be created by any distributed party and change over time, an instance state is a representation captured at one point in time, following from a transition. For tracking, an instance protocol can record models, instances, and instance states over time.
    \item \emph{Blockchain platform}: Distributing models, instances, and instance states can be separated from securing their integrity using a blockchain. For distribution, model repositories can be used, e.g. based on git~\cite{harer_process_2020}. When using cloud-based platforms, they might accomplish distribution, however, with different trade-offs. For securing the integrity of models, hash values are calculated from models, instances, and instance states and recorded with metadata on a blockchain platform. A smart contract implementing corresponding registrations and tracking functions is required.
    \item \emph{Execution engine}: Models must be executed on an execution engine that supports distribution and scalability. For supporting not only models related to organizations, no restrictions are made on execution engines, e.g. including business process management systems and cloud platforms. For the interaction with the blockchain platform, an API is required; possibly implemented on a cloud platform also performing execution.
    \item \emph{Client application}: A client application must support users, e.g. end users and businesses, the interaction with cloud platforms for deploying and running models, as well as the interaction with the blockchain through a suitable API. Interactions encompass controlling the execution in addition to instance tracking by listening to state changes through the blockchain platform such that they become visible to the user. In the context of cloud platforms, web applications may be used together with an API, e.g. web3.js with the MetaMask web browser plugin~\cite{antonopoulosMasteringEthereumBuilding2019}.
\end{enumerate}

\begin{figure}[h]
  \centering
  \includegraphics[clip, trim=0.2cm 0cm 0.1cm 0cm, width=1.02\linewidth]{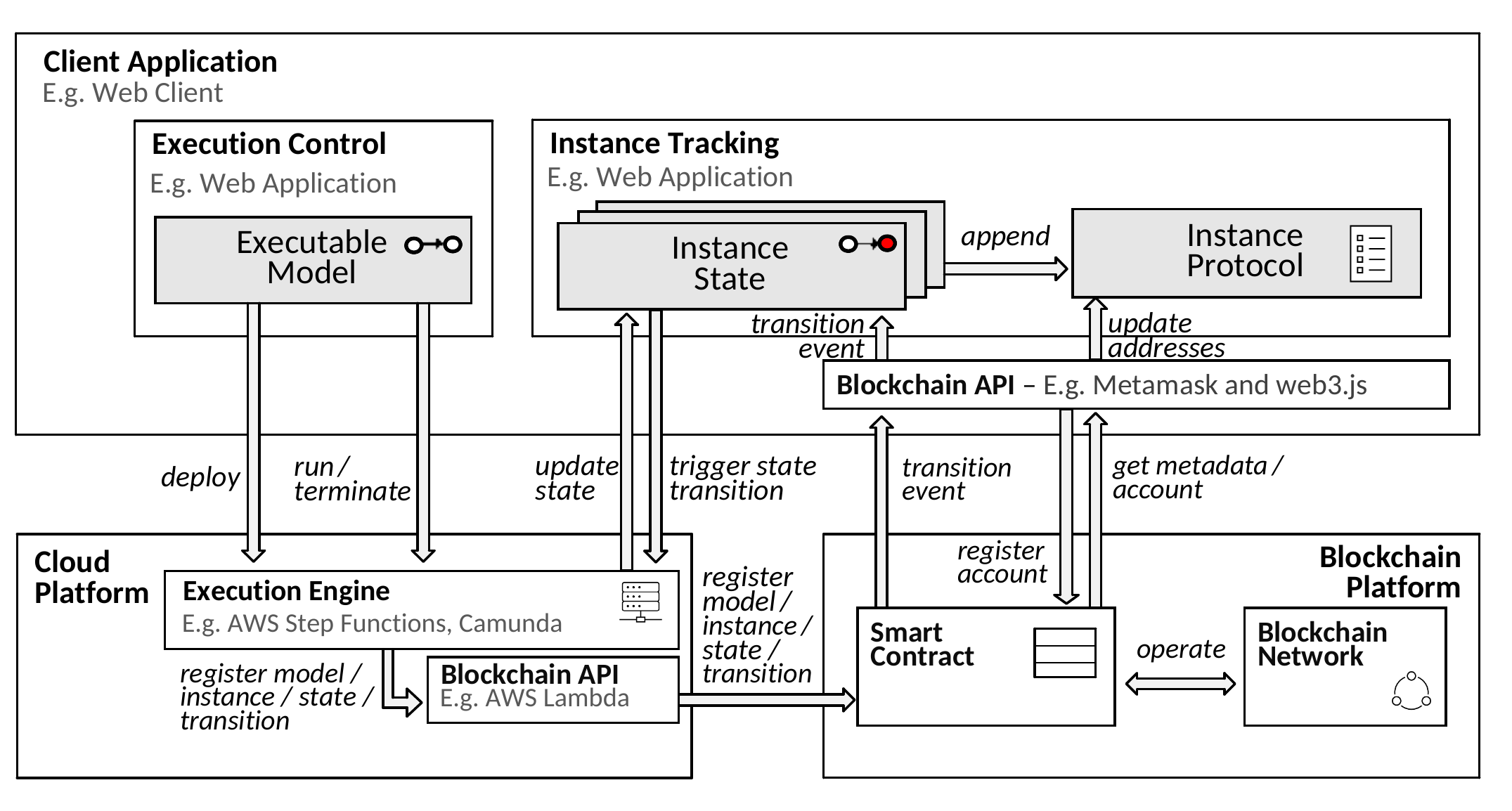}
  \caption{Architecture for executable models with instance tracking.}
  \label{fig:architecture}
\end{figure}

Technically, an architecture fulfilling these requirements can be realized as shown in Figure~\ref{fig:architecture}. 

The client application implements execution control for deploying, running, and terminating an executable model on a cloud platform. Models are registered with the smart contract through the blockchain API component, including metadata and a hash value for each model in order to identify models based on their content. When the execution control component initiates running a model on an execution engine, an instance is registered with the corresponding model hash value and a hash value of the instance.

Changes to the instance are subsequently triggered through the execution engine by user-controlled components or by the engine itself, resulting in new instance states. State changes are captured by the registration of their hash values that reflect and identify new instance states. The pre- and post-state of a transition is thus registered by the corresponding hash values. 

With each transition, the smart contract will emit a transition event on the blockchain network, received by the instance tracking component of each party involved in the execution. There, the event triggers an update of the local state such that state changes can be observed and verified by all parties involved. Each transition and state append the instance protocol, including additional metadata received from the blockchain such as timestamps, owner accounts, or further model-specific data. The result is an instance protocol containing an entry with identifiers for each transition together with the pre- and the post-state, identifiers for the corresponding instance and model, as well as the metadata attribute values.


\subsection{Smart Contract}

\begin{figure}[hb]
  \centering
  \includegraphics[clip, trim=0.4cm 0cm 0.1cm 0.1cm, width=1.02\linewidth]{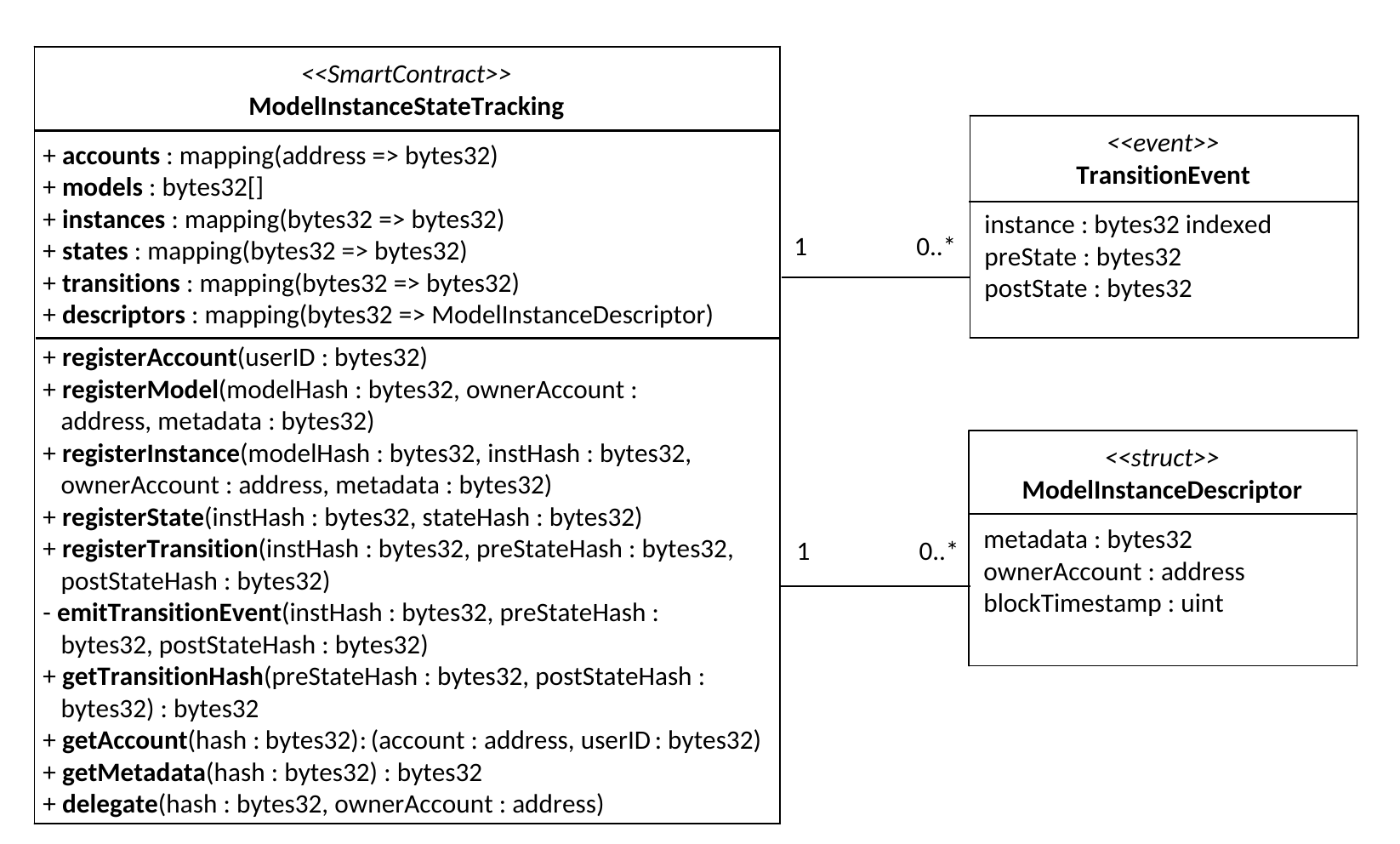}
  \caption{Class diagram of the smart contract for the registering and tracking of models and instances.}
  \label{fig:smartcontract}
\end{figure}

For evaluating implementation feasibility, the smart contract has been implemented as shown in the class diagram in Figure~\ref{fig:smartcontract}. 

Models, instances, states, and transitions are registered through functions interacting with the smart contract component according to the architecture, including the storage by their hash values. Mapping data structures implement storage with additional metadata present in the \texttt{ModelInstanceDescriptor}. In practice, the metadata might depend on the model and modeling tool, e.g. issuing an identifier or name for each model. 

Furthermore, an owner account is stored for each model and instance in order to restrict the creation of instances as well as instance states and transitions to the respective owner. Accordingly, functions for retrieving the account and for delegating access rights exist. 

With each registered transition, the client-side component is notified with a \texttt{TransitionEvent} according to the architecture. The event triggers a client-side update by transmitting the instance with the pre- and the post-state hash values. Given these values, the client-side component might retrieve the corresponding states and transitions together with metadata descriptors. The implementation is available online\footnote{Available from \url{https://github.com/fhaer/Itrex}.}. The smart contract has been deployed and executed on the Ethereum blockchain\footnote{Transactions can be seen at address 0x2300646ca8e90ecb87144432eb975a395e02b7ec, e.g. at \url{https://etherscan.io/address/0x2300646ca8e90ecb87144432eb975a395e02b7ec}.}.

\section{Discussion and Outlook}

For decentralized applications, the research question of coupling executable models with their instances to smart contracts has been initially raised by this paper. As a result, requirements with a corresponding architecture and a smart contract have been constructed through exploratory research. The architecture shows, in principle, executable models can be coupled with smart contracts for a decentralized coordination of the execution when using cloud platforms that offer distribution and scalability independent of individual programs or virtual machines. Serverless computing approaches are well suited for this purpose and allow for data processing involving, e.g., decentralized organizations, processes, tokens, IoT devices, compute-intensive applications such as machine learning, or databases and data-intensive applications.

The feasibility of implementing the architecture with a smart contract is confirmed by a prototype for the Ethereum blockchain, showing the registration of instance states on-chain through hash values and metadata together with an event-based mechanism for updating instance states at the client side. While feasibility could be shown in principle by the architecture, it is a first step towards a model-based approach for executable models in decentralized applications. 

In future research, further analysis of the requirements, implementation, and an extended evaluation are foreseen, aiming at the realization of decentralized coordination through model-based representations.

\begin{acknowledgments}
  This work is partially supported by the Swiss National Science Foundation project Domain-Specific Conceptual Modeling for Distributed Ledger Technologies [196889].
\end{acknowledgments}

\bibliography{references}


\end{document}